# On the Dynamic Statistical Information Theory

## Xing Xiu-San

(Department of Physics, Beijing Institute of Technology, Beijing 100081, China)

E-mail: xiusan@publicb.bta.net.cn

## Abstract

In this paper we extend present Shannon's static statistical information theory to dynamic processes and establish a dynamic statistical information theory, whose theme is the evolution law of dynamic information and dynamic information entropy and its application. Starting from the state variable evolution equation we derive the nonlinear evolution equations of dynamic information density and dynamic information entropy density, that describe respectively the evolution law of dynamic information and dynamic information entropy. These two equations show that the temporal change rate of dynamic information density originates from its drift, diffusion and dissipation in coordinate space and state variable space; and that the temporal change rate of dynamic information entropy density is caused by its drift, diffusion and production in coordinate space and state variable space. Furthermore we present the expressions of drift information flow and diffusion information flow, the formulas of information entropy production rate and information dissipation rate. We prove that the information dissipation rate is equal to the information entropy production rate in a same dynamic system, and that information diffusion and information dissipation occur at the same time. We obtain the dynamic mutual information and dynamic channel capacity reflecting the dynamic dissipation character in the transmission process, which change into their maximum—the present static mutual information and static channel capacity under the limit case when the proportion of channel length to signal transmission rate approaches to zero. All these derivations and results are unified and rigorous from evolution equations of dynamic information and dynamic information entropy without adding any extra assumption. As the application of these theoretical formulas, we investigate and calculate the information dissipation and information entropy production and their time rate for two actual dynamic topics: Brownion motion and molecular motor; and present the dynamic mutual information and dynamic information capacity of a Gaussian channel.

## 1. Introduction

Nowadays Shannon's statistical information theory[1-3] has became an important discipline. The information and the information entropy, as the fundamental concept and quantity in information theory, is used to measure not only the amount of information, but also the degree of order and disorder of a natural or social systems[4]. The concepts and methods of information theory have been applied widely in communication, computer, cybernetics, mathematics, physics, chemistry, biology, economics, sociology, etc[4-8]. Even so, up to now the perfect part of Shannon's statistical information theory only is limited to static state or equilibrium state of time-space-independent process. However, the information and information entropy always change with time and space from the viewpoint of either the information transmission or the degree of order and disorder of dynamic systems. This is the origin of dynamic statistical information theory.

In recent years although information dynamics, information flow and information diffusion have been applied often[9-12], they are, in the main, qualitative and lacking in systematic quantitative theory. Some years ago, the author[13-15] proposed a new fundamental equation of nonequilibrium statistical physics in place of the Liouville equation. That is the anomalous Langevin equation in 6N dimensional phase space or its equivalent Liouville diffusion equation of time-reversal asymmetry. Furthermore we first obtained a nonlinear evolution equation of nonequilibrium physical entropy density changing with time and space, predicted the existence of entropy diffusion, and derived a formula for entropy production rate. On this basis, some questions naturally arise: now that the information and information entropy are the fundamental quantity in information theory and the mathematical expression of information entropy is the same as that of physical entropy, how does dynamic information and dynamic information entropy change with time, space and the other state parameters? Do they obey any evolution equation? What are the forms of these equations if they exist? How to express information entropy production, information dissipation, information flow and information diffusion? What parameters they are related to? Is there any connection between them and the evolution equation? Do the mutual information and channel capacity in dynamic transmission processes need any modification? How to modify? What is the difference between dynamic mutual information, dynamic channel capacity and the present static ones? We shall solve all these problems and present their mathematical expressions from the evolution equations of the information and information entropy in a unified fashion, and use the theoretical formulation to some actual dynamic topics. This is the theme of our dynamic or nonequilibrium statistical information theory. The present static statistical information theory may be regarded as its a special part of time-space-independent process.

## 2. Information entropy evolution equation

Now let us derive the information entropy evolution equation. Because the information entropy is the generalization of physical entropy, and the foundation of physical entropy evolution equation is the new fundamental equation of nonequilibrium statistical physics, we can guess naturally that not only the information entropy evolution equation should exist but also its foundation should be the state variable evolution equation. Let us start from here.

### 2.1 The state variable evolution equation

Suppose that the dynamic system or nonequilibrium system is such as hydrodynamics, grain (crystalline grain, star, etc) growth dynamics and any information symbol change dynamics etc, its state can be expressed by the state variable $a$, such as the velocity or the temperature of fluid, and the size of grain and any information symbol etc. In other words, here the state variables just are information symbols. Let $t$ be the evolution time of state variable, $\dot{a}$ is the change rate of $a$ at time $t$, $\dot{x}$ is the transmission rate of state variable in coordinate space. Because both the dynamic system itself and the state variable as information symbol in transmission processes in coordinate space are always disturbed by internal and external noises, the evolution of state variable changing with time, namely the state variable evolution equation of dynamic system, should be described by the following Langevin equation[8]

$$\begin{cases} \dot{a} = h(a) + \eta_1(t) + \mu(a)\eta_2(t) \\ \dot{x} = v + \varsigma(t) \end{cases} \quad (1)$$

Where $h(a)$ is the drift change rate of state variable itself, which is determined by the property of the dynamic system and generalized force, $\eta_1(t) + \mu(a)\eta_2(t)$ is its fluctuation change rate caused by addictive noise $\eta_1(t)$ and multiplicative noise $\mu(a)\eta_2(t)$, $\mu(a)$ is multiplicative factor of multiplicative noise; $v$ is the drift transmission rate of state variable in coordinate space, $\varsigma(t)$ is its fluctuation change rate. For simplicity, we assume that $\eta_1(t)$, $\eta_2(t)$ and $\varsigma(t)$ are the Gaussian white noise, that is

$$\begin{cases} \langle \eta_1(t) \rangle = \langle \eta_2(t) \rangle = \langle \varsigma(t) \rangle = 0 \\ \langle \eta_1(t)\eta_1(t') \rangle = 2D_1\delta(t-t') \\ \langle \eta_2(t)\eta_2(t') \rangle = 2D_2\delta(t-t') \\ \langle \eta_1(t)\eta_2(t') \rangle = \langle \eta_2(t)\eta_1(t') \rangle = 2\lambda\sqrt{D_1 D_2}\delta(t-t') \\ \langle \varsigma(t)\varsigma(t') \rangle = 2Q\delta(t-t') \end{cases} \quad (2)$$

Where $D_1$, $D_2$ and $Q$ are the noise strength or diffusion coefficient of addictive noise, multiplicative noise and space noise respectively, $\delta(t-t')$ is the Dirac $\delta$ function, $\lambda$ is the correlation strength between $\eta_1(t)$ and $\eta_2(t)$, and $0 \leq \lambda \leq 1$. Make a stochastic transformation[16], then equation (1) can be changed into

$$\begin{cases} \dot{a} = h(a) + g(a)\gamma(t) \\ \dot{x} = v + \varsigma(t) \end{cases} \quad (1a)$$

Where $\gamma(t)$ is the Gaussian white noise, it satisfies

$$\begin{cases} \langle \gamma(t) \rangle = 0 \\ \langle \gamma(t)\gamma(t') \rangle = 2\delta(t-t') \end{cases} \quad (3)$$

The noise strength is written as

$$g(a) = \left[ D_1 + 2\lambda\sqrt{D_1 D_2}\mu(a) + D_2\mu^2(a) \right]^{1/2} \quad (4)$$

According to the stochastic theory, the temporal evolution equation of probability density being equivalent to the Langevin equation (1)(1a) is expressed as the following Fokker-Planck equation[8]

$$\frac{\partial p(a,x,t)}{\partial t} = -\frac{\partial}{\partial a}\left[A(a)p(a,x,t)\right] - v\frac{\partial p(a,x,t)}{\partial x} \qquad (5)$$
$$+ \frac{\partial^2}{\partial a^2}\left[B(a)p(a,x,t)\right] + Q\frac{\partial^2 p(a,x,t)}{\partial x^2}$$

This is the differential equation describing the probability density of the state variable of the dynamic system, where

$$A(a) = h(a) + g(a)\frac{\partial g(a)}{\partial a}, \qquad B(a) = g^2(a) \qquad (6)$$

$p(a,x,t)dadx$ is the probability of finding the state variable i.e. the information symbol between $a$ and $a+da$ within coordinate space $x$ and $x+dx$ at time $t$. Obviously, $p(a,x,t)dadx$ satisfies the normalization condition $\int p(a,x,t)dadx = 1$. The physical meaning of equation (5) is that the temporal change rate of the state variable probability density of the dynamic system originates from its drift and diffusion in state variable space and coordinate space. If A ($a$), B ($a$), $v$ and $Q$ are known, $p(a,x,t)dadx$ can be solved from equation (5). Equationss (1) and (5) are the two equivalent evolution equations of the state variable of dynamic system or the evolution equations of information symbol of dynamic systems.

## 2.2 Information entropy evolution equation

As mentioned at the beginning of this paper, although there are some researches on the temporal change rate of the information entropy, no information entropy evolution equation is available in literature. Now let us derive this equation based on the state variable evolution equation (5).

According to information theory and the correspondence with nonequilium statistical physical entropy, the dynamic information entropy of a dynamic system evolving at time $t$ can be defined as[2,4,13]

$$S(t) = -\int p(a,x,t)\log\frac{p(a,x,t)}{p_m(a,x)}dadx + S_m = \int S_{ax}(t)dadx + S_m \qquad (7)$$

Where $p_m = p_m(a,x)$ is the probability density of maximum information entropy $S_m$ ($p_m$ is called equilibrium probability density in physics) and is a equilibrium solution of the following equation

$$\frac{\partial p_m}{\partial t} = -\frac{\partial}{\partial a}(Ap_m) - v\frac{\partial p_m}{\partial t} + \frac{\partial^2}{\partial a^2}(Bp_m) + Q\frac{\partial^2 p_m}{\partial x^2} = 0 \qquad (8)$$

Where
$$S_{ax}(t) = -p(a,x,t)\log\frac{p(a,x,t)}{p_m(a,x)} \qquad (9)$$

is information entropy density per unit coordinate space and state variable space. Since △S($t$)=S($t$)--S$_m$ is the relative entropy, S$_{ax}$($t$) should be called relative information entropy density in the strict sense.

Differentiating bath sides of equation (7) with respect to time $t$ and using the state variable evolution equations (5) and (8), we obtain the temporal change rate of information entropy

$$\frac{\partial S}{\partial t} = \int \frac{\partial p}{\partial t}(\log \frac{p}{p_m}+1)dadx = \int \frac{\partial S_{ax}}{\partial t}dadx = -\int (\frac{\partial}{\partial a}J_a^s + \frac{\partial}{\partial x}J_x^s - \sigma)dadx \qquad (10)$$

We may write it in the form of the information entropy density

$$\frac{\partial S_{ax}}{\partial t} = -\frac{\partial J_a^s}{\partial a} - \frac{\partial J_x^s}{\partial x} + \sigma \qquad (11)$$

This is the balance equation of information entropy density corresponding to the balance equation of the mass density, momentum density, energy density and physical entropy density. In equations (10) and (11), the quantities are as follows

The information entropy flow density in the state variable space

$$J_a^s = AS_{ax} - \frac{\partial}{\partial a}(BS_{ax}) \qquad (12)$$

The information entropy flow density in the coordinate space

$$J_x^s = vS_{ax} - Q\frac{\partial S_{ax}}{\partial x} \qquad (13)$$

It is obvious that the information entropy flow in both state variable space and coordinate space consist of drift information entropy flow and diffusion information entropy flow.

The information entropy production density

$$\sigma = \sigma_a + \sigma_x = Bp\left(\frac{\partial}{\partial a}\log\frac{p}{p_m}\right)^2 + Qp\left(\frac{\partial}{\partial x}\log\frac{p}{p_m}\right)^2 \qquad (14)$$

Equations (10) and (11) show that the temporal change rate of information entropy in the dynamic system consists of two parts : the information entropy flow and the information entropy production

Substituting formulas (9) (12)—(14) into equation (11), we obtain the evolution equation of dynamic information entropy density $S_{ax}(t)$ as follows

$$\frac{\partial S_{ax}}{\partial t} = -\frac{\partial}{\partial a}(AS_{ax}) - v\frac{\partial S_{ax}}{\partial x} + \frac{\partial^2}{\partial a^2}(BS_{ax}) + Q\frac{\partial^2 S_{ax}}{\partial x^2}$$

$$+\frac{B}{p}\left[\left(\frac{\partial}{\partial a}\log p\right)S_{ax} - \frac{\partial S_{ax}}{\partial a}\right]^2 + \frac{Q}{p}\left[\left(\frac{\partial}{\partial x}\log p\right)S_{ax} - \frac{\partial S_{ax}}{\partial x}\right]^2 \qquad (15)$$

Integrating both sides of equation (15) with respect to $a$ and $x$, we obtain the evolution equations of dynamic information entropy density per unit coordinate space $S_x(t)$ and dynamic information entropy density per unit state variable space $S_a(t)$ as follows

$$\frac{\partial S_x}{\partial t} = -v\frac{\partial S_x}{\partial x} + Q\frac{\partial^2 S_x}{\partial x^2} + \int\frac{B}{p}\left[\left(\frac{\partial}{\partial a}\log p\right)S_{ax} - \frac{\partial S_{ax}}{\partial a}\right]^2 da$$

$$+\int\frac{Q}{p}\left[\left(\frac{\partial}{\partial x}\log p\right)S_{ax} - \frac{\partial S_{ax}}{\partial x}\right]^2 da \qquad (16)$$

$$\frac{\partial S_a}{\partial t} = -\frac{\partial}{\partial a}(AS_a) + \frac{\partial^2}{\partial a^2}(BS_a) + \int \frac{B}{p}\left[\left(\frac{\partial}{\partial a}\log p\right)S_{ax} - \frac{\partial S_{ax}}{\partial a}\right]^2 dx$$

$$+ \int \frac{Q}{p}\left[\left(\frac{\partial}{\partial x}\log p\right)S_{ax} - \frac{\partial S_{ax}}{\partial x}\right]^2 dx \tag{17}$$

Where
$$S_x(t) = \int S_{ax}(t)da = -\int p(a,x,t)\log\frac{p(a,x,t)}{p_m(a,x)}da \tag{18}$$

$$S_a(t) = \int S_{ax}(t)dx = -\int p(a,x,t)\log\frac{p(a,x,t)}{p_m(a,x)}dx \tag{19}$$

In obtaining equations (16) and (17) we have used the boundary conditions that $S_{ax}$ and its derivatives equal to zero at $a \to \pm\infty$ and $x \to \pm\infty$.

Since the equations. (15)(16)(17) include the probability density $p=p(a, x, t)$, they are not closed. Hence, using the expanding formula of the inverse function from formula (9), we obtain

$$p \approx p_m - S_{ax} - \frac{S_{ax}^2}{2p_m}$$

Substituting this approximate formula or the further approximate result $p \approx p_m$ into equations (15)(16)(17), they become closed.

Equations (15)(16)(17) are three evolution equations of dynamic information entropy densities. Among them the equation (15) is basic, and equations (16)(17) are its derivative. They show: the temporary change rate of dynamic information entropy density (the term on the left hand side) is caused together by its drift (the first term on the right hand side), diffusion (the second term on the right-hand side) and production (the third and fourth terms on the right-hand side) in coordinate space and /or state variable space. The information entropy diffusion and production occur at the same time and originate from stochastic noise. Since information entropy measures the degree of disorder of the dynamic systems, the evolution equations (15)(16)(17) represent the nonlinear evolution equations of the density of degree of disorder changing with time, coordinate space and /or state variable space of dynamic systems.

So long as A (*a*), B (*a*), *v* and *Q* are known, $S_{ax}$ (*t*). $S_x$ (*t*) and $S_a$ (*t*) can be solved in principle from the evolution equations (15)(16)(17). But since equations (15)(16)(17) are unclosed nonlinear partial differential equation, it is difficult to get a rigorous solution .

It is not difficult to generalize the above results to multidimensional dynamic systems, whose state variables are $\boldsymbol{a} = (a_1, a_2, \cdots a_n)$ and transmit in three dimensional coordinate space $\boldsymbol{x}=(x_1, x_2, x_3)$. Like equation (5), its probability density $p(\boldsymbol{a},\boldsymbol{x},t)$ obeys multidimensional Fokker-Planck equation, and satisfies the normalization condition $\int p(\boldsymbol{a},\boldsymbol{x},t)d\boldsymbol{a}d\boldsymbol{x}=1$. According to equation (15), the evolution equation of information entropy density of multidimensional dynamic systems should be as follows

$$\frac{\partial S_{ax}}{\partial t} = -\sum_{i}\frac{\partial}{\partial a_i}\left[A_i(\boldsymbol{a})S_{ax}\right] - \sum_{k}v_k\frac{\partial S_{ax}}{\partial x_k} + \sum_{i,j}\frac{\partial}{\partial a_i}\frac{\partial}{\partial a_j}\left[B_{ij}(\boldsymbol{a})S_{ax}\right] + \sum_{k}Q_k\frac{\partial^2 S_{ax}}{\partial x_k^2}$$
$$+\sum_{i,j}\frac{B_{ij}(\boldsymbol{a})}{p(\boldsymbol{a},\boldsymbol{x},t)}\left[\left(\frac{\partial}{\partial a_i}\log p(\boldsymbol{a},\boldsymbol{x},t)\right)S_{ax} - \frac{\partial S_{ax}}{\partial a_i}\right]\left[\left(\frac{\partial}{\partial a_j}\log p(\boldsymbol{a},\boldsymbol{x},t)\right)S_{ax} - \frac{\partial S_{ax}}{\partial a_j}\right]$$
$$+\sum_{k}\frac{Q_k}{p(\boldsymbol{a},\boldsymbol{x},t)}\left[\left(\frac{\partial}{\partial x_k}\log p(\boldsymbol{a},\boldsymbol{x},t)\right)S_{ax} - \frac{\partial S_{ax}}{\partial x_k}\right]^2$$
(20)

Where the information entropy density

$$S_{ax} = S_{ax}(\boldsymbol{a},\boldsymbol{x},t) = -p(\boldsymbol{a},\boldsymbol{x},t)\log\frac{p(\boldsymbol{a},\boldsymbol{x},t)}{p_m(\boldsymbol{a},\boldsymbol{x})} \tag{21}$$

The information entropy production density

$$\sigma_a = \sum_{i,j}B_{ij}(\boldsymbol{a})p(\boldsymbol{a},\boldsymbol{x},t)\left[\frac{\partial}{\partial a_i}\log\frac{p(\boldsymbol{a},\boldsymbol{x},t)}{p_m(\boldsymbol{a},\boldsymbol{x})}\right]\left[\frac{\partial}{\partial a_j}\log\frac{p(\boldsymbol{a},\boldsymbol{x},t)}{p_m(\boldsymbol{a},\boldsymbol{x})}\right]$$
$$+\sum_{k}Q_k p(\boldsymbol{a},\boldsymbol{x},t)\left[\frac{\partial}{\partial x_k}\log\frac{p(\boldsymbol{a},\boldsymbol{x},t)}{p_m(\boldsymbol{a},\boldsymbol{x})}\right]^2 \tag{22}$$

The information entropy flow density

$$J^s_{a_i} = A_i(\boldsymbol{a})S_{ax} - \sum_{j}\frac{\partial}{\partial a_j}\left[B_{ij}(\boldsymbol{a})S_{ax}\right]$$

$$J^s_{x_k} = v_k S_{ax} - Q_k\frac{\partial S_{ax}}{\partial x_k} \tag{23}$$

Where $A_i(\boldsymbol{a})$ is the drift change rate of state variable $\boldsymbol{a}$, $B_{ij}(\boldsymbol{a})$ is its components of diffusion tensor $\boldsymbol{B}$. $v_k$ is the transmission rate component $Q_k$ is its diffusion coefficient.

Obviously, the mathematical form of the evolution equation of information entropy density in $n$ dimensional dynamic systems is same as that one in 1 and 2 dimensional dynamic system.

## 3. Information evolution equation

At first information is applied to measure the amount of information in communication, now also to represent the degree of order of a system. Information density represents the density of degree of order in a system. The dynamic information of dynamic system always changes with time, space and other state variables, its evolution law should be described by a differential equation.

According to information theory, the dynamic information of a dynamic system evolving $t$ time can be defined as[4]

$$I(t) = S_m - S(t) = \int p(\boldsymbol{a},\boldsymbol{x},t)\log\frac{p(\boldsymbol{a},\boldsymbol{x},t)}{p_m(\boldsymbol{a},\boldsymbol{x})}d\boldsymbol{a}d\boldsymbol{x} = \int I_{ax}(t)d\boldsymbol{a}d\boldsymbol{x} \tag{24}$$

Where

$$I_{ax}(t) = p(a,x,t)\log\frac{p(a,x,t)}{p_m(a,x)} \tag{25}$$

is the (relative) information density per unit coordinate space and state variable space. This expression is Kullback information[2,6] in fact.

Equations (24)(7) and (25)(9) show that

$$I(t) + S(t) = S_m \quad \text{and} \quad I_{ax}(t) + S_{ax}(t) = 0 \tag{26}$$

i.e. the sum of information (density) and information entropy (density) in a same dynamic system is constant.

Substituting formula (26) into formulas (25) and equation (15), we obtain the evolution equation of dynamic information density $I_{ax}(t)$ as follows

$$\frac{\partial I_{ax}}{\partial t} = -\frac{\partial}{\partial a}(AI_{ax}) - v\frac{\partial I_{ax}}{\partial x} + \frac{\partial^2}{\partial a^2}(BI_{ax}) + Q\frac{\partial^2 I_{ax}}{\partial x^2}$$

$$-\frac{B}{p}\left[\left(\frac{\partial}{\partial a}\log p\right)I_{ax} - \frac{\partial I_{ax}}{\partial a}\right]^2 - \frac{Q}{p}\left[\left(\frac{\partial}{\partial x}\log p\right)I_{ax} - \frac{\partial I_{ax}}{\partial x}\right]^2 \tag{27}$$

The information dissipation density

$$\rho = \rho_a + \rho_x = -Bp\left(\frac{\partial}{\partial a}\log\frac{p}{p_m}\right)^2 - Qp\left(\frac{\partial}{\partial x}\log\frac{p}{p_m}\right)^2 \tag{28}$$

Integrating both sides of equation (27) with respect to $a$ and $x$, we obtain the evolution equation of dynamic information density per unit coordinate spac $I_x(t)$ and dynamic information density per unit state variable space $I_a(t)$ as follows

$$\frac{\partial I_x}{\partial t} = -v\frac{\partial I_x}{\partial x} + Q\frac{\partial^2 I_x}{\partial x^2} - \int\frac{B}{p}\left[\left(\frac{\partial}{\partial a}\log p\right)I_{ax} - \frac{\partial I_{ax}}{\partial a}\right]^2 da$$

$$-\int\frac{Q}{p}\left[\left(\frac{\partial}{\partial x}\log p\right)I_{ax} - \frac{\partial I_{ax}}{\partial x}\right]^2 da \tag{29}$$

$$\frac{\partial I_a}{\partial t} = -\frac{\partial}{\partial a}(AI_a) + \frac{\partial^2}{\partial a^2}(BI_a) - \int\frac{B}{p}\left[\left(\frac{\partial}{\partial a}\log I_{ax}\right) - \frac{\partial I_{ax}}{\partial a}\right]^2 dx$$

$$-\int\frac{Q}{p}\left[\left(\frac{\partial}{\partial x}\log p\right)I_{ax} - \frac{\partial I_{ax}}{\partial x}\right]^2 dx \tag{30}$$

Where

$$I_x(t) = \int I_{ax} da = \int p(a,x,t)\log\frac{p(a,x,t)}{p_m(a,x)} da \tag{31}$$

$$I_a(t) = \int I_{ax} dx = \int p(a,x,t) \log \frac{p(a,x,t)}{p_m(a,x)} dx \qquad (32)$$

Equations (27)(29)(30) are three evolution equations of dynamic information densities, which we derived. They show: the temporary change rate of dynamic information density of dynamic systems (the term on the left hand-side) is caused together by its drift (the first term on the right hand-side), diffusion (the second term on the right hand-side) and dissipation (the third and fourth terms on the right hand-side) in coordinate space and/or state variable space. Since information measures the degree of order of the dynamic systems, the evolution equations (27)(29)(30) represent the nonlinear equations of the density of degree of order changing with time, corrdinate space and/or state variable space of dynamic systems.

It should be pointed out here, although the mathematical form of equations (29) and (30) is same, their physical meaning is different. Equation (30) describes that dynamic information density per unit state variable space changes with time and state variable, it does not depend on coordinate space and cannot reflect the interchange of information between the system and its environment. However, equation (29) describes that dynamic information density per unit coordinate space changes with time and space, it can reflect the interchange of information between the system and its environment. Equation (27) is the general expression of equations (29)(30), its physical meaning includes both of the two latter's. We can also thus understand the difference of physical meaning among the evolution equations (15)(16)(17) of dynamic information entropy.

If the system has only stochastic diffusion but no drift change in coordinate space, for example, the pure stochastic transmission of information symbol, the pure diffusion movement of small particle, then the evolution equation (29) reduces to

$$\frac{\partial I_x}{\partial t} = Q \frac{\partial^2 I_x}{\partial x^2} - \frac{Q}{p}\left[\left(\frac{\partial}{\partial x}\log p\right) I_x - \frac{\partial I_x}{\partial x}\right]^2 \qquad (33)$$

i.e. the temporal change rate of dynamic information density is only caused by its diffusion and dissipation in coordinate space. Here $p = p(x,t)$. Since the common origin of information diffusion and information dissipation is stochastic change of state variable in coordinate space, whether both of them exist or not is simultaneous. The information dissipation always exists when information diffusion occurs. This is why information evolution equation (33) is different from the mass diffusion equation. Information evolution equation cannot be reduced to pure information diffusion equation[17] only with diffusion term but without dissipation term. The reason is that both mass and energy are conservative, however information and entropy do not.

Obvious, the information entropy density evolution equation (16) should also reduce to the form corresponding to equation (33) in the above case.

It should be also pointed out here, the information density evolution equation (27) can be also written as the information density balance equation corresponding to information entropy density equation (11).

Corresponding to equation (20), the information density evolution equation of multidimensional dynamic system should be as follows

$$\frac{\partial I_{ax}}{\partial t} = -\sum_i \frac{\partial}{\partial a_i}\left[A_i(a) I_{ax}\right] - \sum_k v_k \frac{\partial I_{ax}}{\partial x_k} + \sum_{i,j} \frac{\partial}{\partial a_i}\frac{\partial}{\partial a_j}\left[B_{ij}(a) I_{ax}\right] + \sum_k Q_k \frac{\partial I_{ax}}{\partial x_k^2}$$

$$-\sum_{i,j}\frac{B_{ij}(\boldsymbol{a})}{p(\boldsymbol{a},\boldsymbol{x},t)}\left[\left(\frac{\partial}{\partial a_i}\log p(\boldsymbol{a},\boldsymbol{x},t)\right)I_{ax}-\frac{\partial I_{ax}}{\partial a_i}\right]\left[\left(\frac{\partial}{\partial a_j}\log p(\boldsymbol{a},\boldsymbol{x},t)\right)I_{ax}-\frac{\partial I_{ax}}{\partial a_j}\right]$$

$$-\sum_k \frac{Q_k}{p(\boldsymbol{a},\boldsymbol{x},t)}\left[\left(\frac{\partial}{\partial x_k}\log p(\boldsymbol{a},\boldsymbol{x},t)\right)I_{ax}-\frac{\partial I_{ax}}{\partial x_k}\right]^2 \quad (34)$$

where the information density

$$I_{ax}=I_{ax}(\boldsymbol{a},\boldsymbol{x},t)=p(\boldsymbol{a},\boldsymbol{x},t)\log\frac{p(\boldsymbol{a},\boldsymbol{x},t)}{p_m(\boldsymbol{a},\boldsymbol{x})} \quad (35)$$

the information dissipation density

$$\rho_a = -\sum_{i,j}B_{ij}(\boldsymbol{a})p(\boldsymbol{a},t)\left[\frac{\partial}{\partial a_i}\log\frac{p(\boldsymbol{a},t)}{p_m(\boldsymbol{a})}\right]\left[\frac{\partial}{\partial a_j}\log\frac{p(\boldsymbol{a},t)}{p_m(\boldsymbol{a})}\right]$$
$$-\sum_k Q_k p(\boldsymbol{a},\boldsymbol{x},t)\left[\frac{\partial}{\partial x_k}\log\frac{p(\boldsymbol{a},\boldsymbol{x},t)}{p_m(\boldsymbol{a},\boldsymbol{x})}\right]^2 \quad (36)$$

Due to the existence of diffusion and dissipation or production, the information and information entropy evolution equations (27)(29)(30)(34) and (15)(16)(17)(20) are time reversal asymmetrical, which reflect the irreversibility of these evolution processes.

It is obvious from evolution equations (27)(29)(30)(34) and (15)(16)(17)(20), if the system is in static state or equilibrium state, its information density and information entropy density do not change with time $t$ and coordinate space $x$, the information (entropy) flow and information (entropy) dissipation (production) equal to zero. This is just the circumstance of the present static statistical information theory. Hence it can be concluded that the present static statistical information theory may be regarded as a special part of time-space-independent process of dynamic statistical information theory.

If the dynamic system is isolated, it does not exchange information with its environment. Integrating both sides of equation (30) with respect to state variable $a$ and time $t$, we obtain that the dynamic information after evolving (i.e. relaxing) $\tau$ time due to cumulative dissipation changes to

$$I(\tau)=I(0)-\int_0^\tau dt\int\left\{\frac{B}{p}\left[\left(\frac{\partial}{\partial a}\log p\right)I_{ax}-\frac{\partial I_{ax}}{\partial a}\right]^2+\frac{Q}{p}\left[\left(\frac{\partial}{\partial x}\log p\right)I_{ax}-\frac{\partial I_{ax}}{\partial x}\right]^2\right\}dadx$$
(37)

Where the term on the left hand-side is the information of system at time $\tau$, the first term on the right hand-side is the information of system at time $t=0$ the second term on the right hand-side is the cumulative dissipation information of system evolving from time $t=0$ to $t=\tau$.

If the dynamic system is stationary, then from equation (29) $\frac{\partial I_x}{\partial t}=0$, we obtain that the dynamic information transmitting through channel from input terminal $x=0$ to output terminal $x=l$ due to cumulative dissipation in the transmission process changes to

$$I_x(l) = I_x(0) - \frac{Q}{v}\left[\left(\frac{\partial I_x}{\partial x}\right)_0 - \left(\frac{\partial I_x}{\partial x}\right)_l\right] - \frac{1}{v}\int_0^l dx \int da$$

$$\left\{\frac{B}{p_{st}}\left[\left(\frac{\partial}{\partial a}\log p_{st}\right)I_{ax} - \frac{\partial I_{ax}}{\partial a}\right]^2 + \frac{Q}{p_{st}}\left[\left(\frac{\partial}{\partial x}\log p_{st}\right)I_{ax} - \frac{\partial I_{ax}}{\partial x}\right]^2\right\} \tag{38}$$

where $l$ is the channel length, $p_{st} = p_{st}(a,x)$ is probability density of stationary system. The term on the left hand side is the information density at output terminal $x=l$, the first term on the right hand-side is the information density at input terminal $x=0$, the second term on the right hand-side is the information density difference due to the difference of information density gradient at both channel terminals, the third term on the right hand-side is the cumulative dissipation of information density in the transmission process.

## 4. Information flow and information diffusion

It is easy to know from information evolution equation (27) or (29)(30) that the information flow in both state variable space and coordinate space consist of drift information flow and diffusion information flow.

The information flow density in state variable space

$$J_a^{ax} = J_{ta}^{ax} + J_{da}^{ax} = AI_{ax} - \frac{\partial}{\partial a}(BI_{ax}) \tag{39a}$$

or

$$J_a = J_{ta} + J_{da} = AI_a - \frac{\partial}{\partial a}(BI_a) \tag{39b}$$

This information flow only changes the internal information density of a system and does not relate to exchange of information with the environment.

The information flow density in coordinate space

$$J_x^{ax} = J_{tx}^{ax} + J_{dx}^{ax} = vI_{ax} - Q\frac{\partial I_{ax}}{\partial x} \tag{40a}$$

or

$$J_x = J_{tx} + J_{dx} = vI_x - Q\frac{\partial I_x}{\partial x} \tag{40b}$$

This information flow has practical meaning whether in communication process or for investigating the change of the degree of order in open system. This is because the information exchange of a system with its environment (including one system with the other, among respective part in biologic organism, among respective member of society etc) is accomplished by forward and backward transmission of information flow in coordinate space.

The drift information flow density in coordinate space

$$J_{tx}^{ax} = vI_{ax} \tag{41a}$$

or

$$J_{tx} = vI_x \tag{41b}$$

It equals to the product of drift change rate $v$ of information and information density $I_{ax}$ or $I_x$ in coordinate space.

As motioned above, $v$ is in fact the drift change rate of state variable in coordinate space. In communication, when man receives the information with eyes, $v$ is light velocity; when man receives the information with ears, $v$ is sound velocity.

The diffusion information flow density in coordinate space

$$J_{dx}^{ax} = -Q\frac{\partial I_{ax}}{\partial x} \qquad (42a)$$

or

$$J_{dx} = -Q\frac{\partial I_x}{\partial x} \qquad (42b)$$

It equals to the product of diffusion coefficient $Q$ and negative gradient of information density in coordinate space. In other words, information diffusion means that information always diffuse spontaneously from high density region to low density region. In three spaces, the diffusion direction may be isotropic all around, hence it is obviously different from directional transmission of the drift information flow. The rumour transmission is a typical information diffusion. The smell communication of animal is in fact a communication with information diffusion. Due to its importance, the information diffusion in internet has monograph[12]. Just because the information can diffuse spontaneously, to keep the secret of politics, military and finace have been paid attention to extremely by all countries.

Owing to diffusion, the information density and information entropy density of an isolated system will become homogeneous. At final, the system approaches to equilibrium.

It is easy to know from equation (34) that corresponding to formula (23) the information flow density of $n$ dimensional dynamic system is

$$J_{a_i} = A_i(\boldsymbol{a})I_a - \sum_j \frac{\partial}{\partial a_j}\left[B_{ij}(\boldsymbol{a})I_a\right] \qquad (43)$$

Just as equation (38) pointed out that the information density will change in transmission process, the information flow density also decreases with the increase of transmission distance in space owing to cumulative dissipation. If the dynamic system is nonequilibrium, we can obtained the space information flow density at time $t$ through position $x$ from equations (40b)(31) as follows

$$J_x(t) = \int \left\{\left[vp(a,x,t) - Q\frac{\partial p(a,x,t)}{\partial x}\right]\ln\frac{p(a,x,t)}{p_m(a,x)} - Q\left[\frac{\partial p(a,x,t)}{\partial x} - \frac{p(a,x,t)}{p_m(a,x)}\frac{\partial p_m(a,x)}{\partial x}\right]\right\}da$$

(44)

It can be seen that the information flow density is a function of time and space. Thus we can easily obtain the difference of information flow density between both at time $t=0$ through position $x=0$ and at time $t=\tau$ through position $x=l$

$$J_0(0) - J_l(\tau) = P_l(\tau) \qquad (45)$$

This is the change expression of information flow density decreasing with the increase of transmission distance in nonequilibrium transmission process. Where $P_l(\tau)$ is the cumulative dissipation of information flow density in transmission process, it can be calculated from formula (44).

If the dynamic system is stationary, then from equation (29) $\frac{\partial I_x}{\partial t} = -\frac{\partial J_x{}'}{\partial t} = 0$ (here $J_x{}' \neq J_x$) and substituting into formula (40b), we obtain the change expression of information flow density decreasing with the increase of trensmission distance owing to cumulative dissipation in trensmission process

$$J_x = J_0 - \int_0^x dx \int_{-\infty}^{\infty} da \left\{ \frac{B}{p_{st}} \left[ \left( \frac{\partial}{\partial a} \log p_{st} \right) I_{ax} - \frac{\partial I_{ax}}{\partial a} \right]^2 + \frac{Q}{p_{st}} \left[ \left( \frac{\partial}{\partial x} \log p_{st} \right) I_{st} - \frac{\partial I_{ax}}{\partial x} \right]^2 \right\}$$

$$= J_0 - P_x \qquad (46)$$

Where $J_0$ and $J_x$ are the information flow density through position $x=0$ and $x=x$, $P_x$ is cumulative dissipation. They are all independent on time because the system is stationary.

## 5. Information entropy production and information dissipation

Information entropy production and information dissipation are the intrinsic property of complex dynamic systems. Which parameters and how it change with? Can they be described by a quantitative formulas? This is an important problem to be solved in dynamic statistical information theory. Now let us present the concise formulas for information entropy production and information dissipation in this section.

At first, let us define a new parameter of the dynamic system, that is the percentage departure from equilibrium of probability density of the dynamic system as

$$\theta = \log \frac{p}{p_m} \approx \frac{\Delta p}{p_m} \qquad (47)$$

It can be also called the percentage departure from equilibrium for brevity. The gradients of the percentage departure from equilibrium in state variable space and coordinate space are

$$\begin{cases} \nabla_a \theta = \nabla_a \log \frac{p}{p_m} \\ \nabla_x \theta = \nabla_x \log \frac{p}{p_m} \end{cases} \qquad (48)$$

Substituting formula (48) into (14), we obtain the expression for information entropy production rate of the dynamic systems

$$\frac{d_i S}{dt} = \int (\sigma_a + \sigma_x) da dx = \int p \left[ B \left( \frac{\partial \theta}{\partial a} \right)^2 + Q \left( \frac{\partial \theta}{\partial x} \right)^2 \right] da dx$$

$$= \overline{B \left( \frac{\partial \theta}{\partial a} \right)^2} + \overline{Q \left( \frac{\partial \theta}{\partial x} \right)^2} \geq 0 \qquad (49)$$

This is just a concise formula for information entropy production rate, i.e. a concise formula for the law of information entropy increase. Comparing with the law of physical entropy increase, the law of information entropy increase, also named the law of generalized entropy increase, it is useful for not only natural science but also social science. It shows that the information entropy of natural and social dynamic systems always tends to increase, that is, the systems evolve spontaneously to the direction of information entropy increase. Because the information entropy represents the degree of disorder of the dynamic systems, hence the law of information entropy increases means that the degree of disorder of natural and social dynamic systems always tend to increase.

Similarly substituting formula (47)(48) into (28), we obtain the expression for information dissipation rate of the dynamic system

$$\frac{d_i I}{dt} = -\int p \left[ B \left(\frac{\partial \theta}{\partial a}\right)^2 + Q \left(\frac{\partial \theta}{\partial x}\right)^2 \right] dadx = -\left[ \overline{B \left(\frac{\partial \theta}{\partial a}\right)^2} + \overline{Q \left(\frac{\partial \theta}{\partial x}\right)^2} \right] \leq 0 \quad (50)$$

This is just a concise formula for information dissipation rate. Corresponding to the law of information entropy increase, this also may be regarded as a concise formula for the law of information dissipation. It shows that the information of natural and social dynamic systems always tend to dissipate, that is, the systems evolve spontaneously to the direction of information dissipation. Because the information represents the degree of order, hence the law of information dissipation means that the degree of order of natural and social dynamic systems always tend to dissipate.

It can be seen from formulas(49)(50)that both information entropy production rate and information dissipation rate consist of two part: one originates from the interior of dynamic systems(in state variable space), the other occurs in transmission processes( in coordinate space). Both of them are independent.

If the dynamic systems has only stochastic change in coordinate space, then the formulas for information entropy production rate and information dissipation rate corresponding equation (33) are

$$\frac{d_i S}{dt} = \overline{Q \left(\frac{\partial \theta}{\partial x}\right)^2} \geq 0 \quad (51)$$

$$\frac{d_i I}{dt} = -\overline{Q \left(\frac{\partial \theta}{\partial x}\right)^2} \leq 0 \quad (52)$$

Similarly, corresponding to equations (20)(34), the formulas for information entropy production rate and information dissipation rate of $n$ dimensional dynamic systems are

$$\frac{d_i S}{dt} = \int p(\boldsymbol{a},t) B(\boldsymbol{a}) : (\nabla_a \theta)(\nabla_a) da_1 da_2 \cdots da_n$$
$$= \overline{B:(\nabla_a \theta)(\nabla_a \theta)} \geq 0 \quad (53)$$

$$\frac{d_i I}{dt} = -\int p(\boldsymbol{a},t) B(\boldsymbol{a}) : (\nabla_a \theta)(\nabla_a \theta) da_1 da_2 \cdots da_n$$
$$= -\overline{B:(\nabla_a \theta)(\nabla_a \theta)} \leq 0 \quad (54)$$

Formulas (49)—(54) prove, both information entropy production rate and information dissipation rate equal to the average value of the product of diffusion coefficient or noise strength and the square of gradient of the percentage departure from equilibrium. They show that the information entropy production and information dissipation of the dynamic systems are caused by stochastic and inhomogeneous departure from equilibrium of probability density in state variable space and coordinate space. It can be seen, for nonequilibrium($\theta \neq 0$) inhomogeneous ($\partial \theta/\partial a \neq 0$, $\partial \theta/\partial x \neq 0$) complex dynamic systems with stochastic diffusion ($B \neq 0$, $Q \neq 0$), the information entropy always produce, the information always dissipate. Conversely for equilibrium systems (θ=0) or that nonequilibrium but homogeneous (∂θ/∂a=0, ∂θ/∂x=0) dynamic systems or that dynamic systems only with deterministic but no stochastic motion(B=0,Q=0), there are all no both information entropy production and information dissipation.

It should be pointed out, although the mathematical form and physical meaning of information entropy evolution equations (15)(16)(17)(20) of complex dynamic systems are similarly to physical entropy evolution equation[13-15] of statistical thermodynamic systems: the temporary change rate of entropy density is caused by drift, diffusion and production. However, both of them have important difference: the gradient of diffusion term and production term [i.e. the expressions of entropy production rate (14)(22)(49)(53)] in the former can take with respect to both position and velocity or other parameters; the gradient in the latter only take with position, but not velocity or other parameters [13-15].

If the system is stationary, through its macro-state does not change with time, a macroscopic current exists within the system. The information entropy production rate of one dimensional stationary system can be derived from formula (49) and $\frac{\partial p_{st}(a)}{\partial t} = -\frac{\partial J}{\partial a} = 0$ as follows[15]

$$\frac{d_i S}{dt} = J \left[ -\phi(L) + \ln \frac{p_{st}(0)}{p_{st}(L)} \right] \tag{55}$$

Where $J$ is the probability current, $\phi(L) = -\frac{1}{B} \int_0^L h(a) da$, $L$ is the interval length $[O, L]$, in which the $J$ flows. $p_{st}(0)$ and $p_{st}(L)$ are the probability density at boundaries $a=0$ and $a=L$.

Similarly, we can obtain from formulas (50)(55) the information dissipation rate of one dimensional stationary system

$$\frac{d_i I}{dt} = -J \left[ -\phi(L) + \ln \frac{p_{st}(0)}{p_{st}(L)} \right] \tag{56}$$

Formulas (55)(56) show that the information entropy production rate and the information dissipation rate in the stationary state are proportional to their probability current. When the system is in equilibrium state, $J=0$, both the information entropy production rate and the information dissipation rate equal to zero. This agrees with the result of formulas (49)—(54).

It can be seen from (49)—(56) that in a same dynamic system, information dissipates when information entropy produces, information increases when information entropy decreases. Both of them exist simultaneously and their values are equal. In other words, the information entropy production (or decrease) rate is equal to the information dissipation (increase) rate. That is, the temporal change rate of the sum of information entropy and information within a same dynamic system is zero

$$\frac{d_i S}{dt} = -\frac{d_i I}{dt} \quad \text{or} \quad \frac{d_i}{dt}(S+I) = 0 \tag{57}$$

Starting from the laws of information entropy increase and information dissipation it can be concluded that any natural and social systems cannot be as white as a lily and a monolithic bloc, within which there always are disorder subsystems. For example, there are defects in crystal, dead calls in good health organization, criminals in society etc.

Owing to the inhomogeneity of percentage departure from equilibrium of information entropy density and information density, it can be drawn a new inference: the change of corresponding microstructure within the dynamic systems during an irreversible process is inhomogeneous. For example, the dislocation slip always concentrate into slip bands during the plastically deformed process of monocrystal; the ageing and pathologic change speed of all organ are different during the ageing process of man, such as some one has only a serious heart disease, the other has merely a dangerous disease of the kidneys.

## 6. Dynamic mutual information and dynamic channel capacity

A fundamental question in Shannon information transmission theory is to give a formula for calculating the amount of information, which is transmitted through channel by transmitter from the input terminal and received by receiver at the output terminal. This is the present static mutual information[1-3].

$$I(A;B) = H(A) - H(A|B) = H(B) - H(B|A) \tag{58}$$

Where A is the symbol transmitted through channel by transmitter from input terminal, B is the symbol received by receiver at the output terminal. $H(A)$ is the entropy of symbol A, $H(B)$ is the entropy of symbol B, $H(A|B)$ and $H(B|A)$ are the noise entropies caused by noise disturbance on the symbols. It should be pointed out here, information transmission is just that information flow transmits from one point to another point in time-space and is a dynamic process changing with time and same space. If transmitting rate of symbol is $v$, then the transmitting time through the channel length $l$ is $\tau=l/v$. Hence we must mark a time-space coordinate of channel on the symbols, for example, let $A_x(t)$ and $B_x(t)$ be the symbols A and B at time $t$ in the position $x$ of channel. The reason why the symbols of all terms in formula (58) have no obvious time-space coordinate is that the present static communication model is a point model (or zero transmitting time model) in essence. There is no time-space process. That is, that the symbols transmit from the input terminal to the output terminal does not need any time. Both the terminals are appropriately located at same point, the noise is also concentrated on this point. Therefore, the symbols of all terms in formula (58) take the same time and same space coordinate. For mutual information (58), it should be written $I(A;B) = I(A_0(0); B_0(0)) = I(A_l(\tau); B_l(\tau))$, its exact definition is: the receiver at the time $t = \tau(=0)$ and the output terminal $x = l(=0)$ receive the amount of information contained in symbol $B_l(\tau)(= B_0(0))$ about symbol $A_l(\tau)(= A_0(0))$ transmitted by transmitter at a same time and same position. Now the question arise: when the transmitting time of symbols through channel is long enough and there is noise in transmission process, how to calculate that the receiver at the time $t = \tau$ and the output terminal $x = l$ receive the amount of information contained in symbol $B_l(\tau)$ about symbol $A_0(0)$ transmitted by transmitter at the time $t = 0$ and the input terminal $x = 0$? This is dynamic mutual information.

In order to obtain dynamic mutual information, we write static mutual information (58) as follows

$$I(A_l(\tau); B_l(\tau)) = H(A_l(\tau)) - H(A_l(\tau)|B_l(\tau)) = H(B_l(\tau)) - H(B_l(\tau)|A_l(\tau)) \tag{58a}$$

Compared with (58), this formula obviously include transmitting time $\tau$ and the output terminal

$l$, its exact definition has motioned as above. Because dynamic information exists cumulative dissipation in transmission process, the information $H(A_o(0))$ transmitted from the input terminal will reduce to $H(A_l(\tau))$ when it reaches the output terminal by information flow. Their relation is

$$H(A_l(\tau)) = H(A_0(0)) - H(Q_l(\tau)) \tag{59}$$

where $H(Q_l(\tau))$ is the cumulative dissipation of information transmitting from the input terminal to the output terminal. Substituting formula (59) into (58a), we obtain the formula for dynamic mutual information

$$I(A_0(0); B_l(\tau)) = H(A_0(0)) - H(A_0(0)|B_l(\tau)) \tag{60}$$

where

$$H(A_0(0) | B_l(\tau)) = H(Q_l(\tau)) + H(A_l(\tau) | B_l(\tau)) \tag{61}$$

Though formula (60) comes from (58a), its physical meaning is different. This dynamic mutual information just is that the receiver at the time $t=\tau$ and the output terminal $x=l$ receive the amount of information contained in symbol $B_l(\tau)$ about symbol $A_0(0)$ transmitted by transmitter at the time $t=0$ and the input terminal $x=0$. Why formula (60) is called dynamic mutual information? The reason is that it includes dynamic information dissipation in transmission process, but formulas (58)(58a) do not.

In order that formula (60) can be applied to actual calculation, we must know the cumulative information dissipation in transmission process. Substituting the change expression of information flow density in transmission process (45) or (46) into (59), it is not difficult to obtain

$$\begin{cases} H(Q_l(\tau)) = \dfrac{P_l(\tau)}{J_0(0)} H(A_0(0)) \\ H(A_l(\tau)) = \left[1 - \dfrac{P_l(\tau)}{J_0(0)}\right] H(A_0(0)) = \dfrac{J_l(\tau)}{J_0(0)} H(A_0(0)) \end{cases} \tag{62}$$

Substituting formulas (61)(62) into (60), then dynamic mutual information reduces to

$$I(A_0(0); B_l(\tau)) = \dfrac{J_l(\tau)}{J_0(0)} H(A_0(0)) - H(A_l(\tau)|B_l(\tau)) \tag{63}$$

Because formulas (45) and (46) are suitable for both nonequilibrium state and stationary state respectively, dynamic mutual information (63) can be applicable to both nonequilibrium state and stationary state.

The another important question in information transmission theory is the channel capacity, i.e. the largest amount of information that can be transmitted through the channel. In Shannon information theory, the static channel capacity is defined as[1-3]

$$C = \max_{p(a)} I(A; B) \tag{64}$$

Where $I(A;B)$ is the static mutual information (58), $p(a)$ is the input probability density of symbol $A$. Along with the appearance of dynamic mutual information (60)(63), there should be also the dynamic channel capacity. Corresponding to formula (64), it should be defined as

$$C_d = \max_{p(a_0)} I\left(A_0(0); B_l(\tau)\right)$$

$$= \max_{p(a_0)} \left[ \frac{J_l(\tau)}{J_0(0)} H\left(A_0(0)\right) - H\left(A_l(\tau) \mid B_l(\tau)\right) \right] \quad (65)$$

Where $p(a_0)$ is probability density of symbol $A_0(0)$ at the time $t=0$ and the input terminal $x=0$. Formula (65) just comes from formula (64) by means of dynamic mutual information (63) in place of static mutual information (58).

It can be seen from formulas (60)(61)(63)(65) that the information dissipation caused by noise in dynamic mutual information and dynamic channel capacity consist of two parts: the first is the cumulative information dissipation $H(Q_l(\tau))$ in transmission process from the input terminal to the output terminal, which reduces $H(A_0(0))$ to $H(A_l(\tau))$; the second is the instantaneous information dissipation $H(A_l(\tau) \mid B_l(\tau))$ in formulas (58)(58a)(64) occuring at receiving time and the output terminal. Under what condition $H(Q_l(\tau))$ can be neglected? It can be seen from formulas (62)(63) especially the following expressions (76)(84) that the cumulative information dissipation $H(Q_l(\tau))$ can be neglected only when transmitting time $\tau=l/v \to 0$ i.e. the proportion of channel length $l$ to transmitting rate $v$ tends to zero. Under this condition the information transmitting from the input terminal to the output terminal does not take place any change nearly, formulas (60)(63)(65) reduce to formulas (58)(58a)(64). This is just the suitable condition of static mutual information and static channel capacity (58)(58a)(64). Conversely, if channel length $l$ is very long, or transmitting rate $v$ is very small, or noise strength $Q$ is very large, the cumulative information dissipation $H(Q_l(\tau))$ cannot be neglected. Under such circumstance, mutual information represent that the receiver at the output terminal receive the amount of information transmitted from the input terminal and channel capacity should be dynamic formulas (60)(63)(65) in place of static formulas (58)(58a)(63). For examples, the communication in deep ocean or between stars, in the former case the sound transmitting rate is smaller, in the latter case the distance is very long, and the noise is strong in both transmission process, so dynamic mutual information (60)(63) and dynamic channel capacity (65) have its practical mean.

It can be seen that dynamic mutual information and dynamic channel capacity are in general smaller than static mutual information and static channel capacity, only their maximum tend to the latter.

It should pointed out here, in order to correspond entirely to the present static information theory, dynamic information entropy (7) and dynamic information (24) of dynamic system evolving at time $t$ should be defined as[1-3]

$$S(t) = -\int p(a,x,t) \log p(a,x,t) \, dadx = \int S_{ax}(t) \, dadx \quad (7a)$$

$$I(t) = S_m - S(t) = S_m + \int p(a,x,t) \log p(a,x,t) \, dadx = S_m + \int I_{ax}(t) \, dadx$$

(24a)

Although formulas (7a)(24a) and (7)(24) are slightly different from each other in forms, both of them can be applicable. The more importance is when $A, B, v$ and $Q$ are approximately constant ($v$ and $Q$ can be regarded as constant in communication process), the evolution equations of information entropy density $S_{ax}(t)$ in (7a) and information density $I_{ax}(t)$ in (24a) are the same as equations (15) and (27) in forms. The difference is that corresponding with expressions (14) and (28) the information entropy production density and information dissipation density here are

$$\sigma = Bp\left(\frac{\partial}{\partial a} \log p\right)^2 + Qp\left(\frac{\partial}{\partial x} \log p\right)^2$$

(14a)

$$\rho = -Bp\left(\frac{\partial}{\partial a} \log p\right)^2 - Qp\left(\frac{\partial}{\partial x} \log p\right)^2$$

(28a)

If the dynamic system is nonequilibrium, corresponding to expression (44), the information flow density at time $t$ through position $x$ here are

$$J_x(t) = \int \left\{ \left[vp(a,x,t) - Q\frac{\partial p(a,x,t)}{\partial x}\right] \ln p(a,x,t) - Q\frac{\partial p(a,x,t)}{\partial x} \right\} da$$

(44a)

Compared with (14a)(28a), the superiority of expressions (14)(28) is that the information entropy production and information dissipation are equal to zero automatically at equilibrium state $p=p_m$, it need not any additive condition. However, if we want that the information entropy production and information dissipation described by formulas (14a)(28a) equal to zero at equilibrium state, it must need $\left(\frac{\partial p}{\partial a}\right)_{p_m} = \left(\frac{\partial p}{\partial x}\right)_{p_m} = 0$. In general, the former is not reasonable.

This is why expressions (7) and (14) are applied as the definition of dynamic information entropy and dynamic information in this paper.

Nowadays some men try to derive all physical fundamental equations in a unified fashion from the maximum principle of Fisher information[18]. The information entropy production rate expressed by formula (14a) during $B=0$ and $Q=1$ just equals to Fisher information. That is to say, Fisher information is a special example of Shannon information entropy production rate.

### 7. Application

Now let us apply the above theoretical formulation in sections (2)—(6) to investigate some actual topics.

### 7.1 Brownian motion

Suppose the state variable itself of system takes place Brownian motion and its position transmits with drift-diffusion in coordinate space at the same time. For example, the dynamic Gaussian signal transmitted through spatial Gaussian channel. In this transmission process, information entropy and information will change. According to equation (5), the Fokker-Planck equation describing this dynamic system is[19]

$$\frac{\partial p}{\partial t} = \beta \frac{\partial}{\partial a}(ap) - v\frac{\partial p}{\partial x} + D\frac{\partial^2 p}{\partial a^2} + Q\frac{\partial^2 p}{\partial x^2} \tag{66}$$

where β is friction constant. If the initial solution of equation (66) is

$$p(a,x,t=0) = (\pi\ \varepsilon_0)^{-1/2} \exp\left[-\frac{(a-a_0)^2}{\varepsilon_0}\right] \times 2(\pi\varphi)^{-1/2} \exp\left(-\frac{x^2}{\varphi}\right) \tag{67}$$

then its time-dependent solution can be obtained[19]

$$p(a,x,t) = (\pi\ \varepsilon_t)^{-1/2} \exp\left\{-\frac{[a-b(t)]^2}{\varepsilon_t}\right\} \times 2(\pi\varphi+4\pi Qt)^{-1/2} \exp\left[-\frac{(x-vt)^2}{(\varphi+4Qt)}\right]$$

$$= p(a,t)\,p(x,t) \tag{68}$$

Where
$$\begin{cases} \varepsilon_t = \varepsilon_0 e^{-2\beta t} + \varepsilon_m\left(1-e^{-2\beta t}\right) \\ b(t) = b_0 e^{-\beta t}, \varepsilon_m = 2D/\beta \end{cases} \tag{69}$$

The equilibrium solution of equation (66) is

$$p_m(a,x) = p_m(a)\,p_m(x), \quad \begin{cases} p_m(a) = (\pi\varepsilon_m)^{-1/2} \exp\left(-\frac{a^2}{\varepsilon_m}\right) \\ p_m(x) = constant \end{cases} \tag{70}$$

Substituting formulas (68)(70) into (48)(49), we obtain the expression of the information entropy production rate of this system at time $t$

$$P(t) = \frac{d_i S}{dt} = \int_{-\infty}^{\infty} da \int_0^{\infty} dx \left[ Dp\left(\frac{\partial}{\partial a}\log\frac{p}{p_m}\right)^2 + QP\left(\frac{\partial}{\partial x}\log\frac{p}{p_m}\right)^2 \right]$$

$$= \beta\left[\frac{2b^2(t)}{\varepsilon_m} + \frac{\varepsilon_t}{\varepsilon_m} + \frac{\varepsilon_m}{\varepsilon_t} - 2\right] + \frac{2Q}{4Qt+\varphi} \geq 0 \tag{71}$$

The temporal change rate of information entropy production rate at time $t$ is

$$\frac{\partial P(t)}{\partial t} = -2\beta^2 \left[\frac{2b^2(t)}{\varepsilon_m} + \frac{\varepsilon_t}{\varepsilon_m} + \frac{\varepsilon_m^2}{\varepsilon_t^2} - \frac{\varepsilon_m}{\varepsilon_t} - 1\right] - \frac{8Q^2}{(4Qt+\varphi)^2} \leq 0 \tag{72}$$

The information entropy production of this system at time $t$ is

$$\Delta_i S = \int_0^t P(t)\,dt = \frac{b_0^2}{\varepsilon_m}\left(1-e^{-2\beta t}\right) + \frac{1}{2}\left(e^{-2\beta t}-1\right) + \frac{\varepsilon_0}{2\varepsilon_m}\left(1-e^{-2\beta t}\right)$$

$$+ \frac{1}{2}\ln\frac{\varepsilon_m+(\varepsilon_0-\varepsilon_m)e^{-2\beta t}}{\varepsilon_0} + \frac{1}{2}\ln\frac{4Qt+\varphi}{\varphi} \geq 0 \tag{73}$$

According to formulas (50)(57)(68)—(73), we can obtain the information dissipation rate, its temporal change rate and the information dissipation of this dynamic system at time $t$

$$\begin{cases} P_i = \dfrac{d_i I}{dt} = -\dfrac{d_i S}{dt} = -\beta \left[ \dfrac{2b^2(t)}{\varepsilon_m} + \dfrac{\varepsilon_t}{\varepsilon_m} + \dfrac{\varepsilon_m}{\varepsilon_t} - 2 \right] - \dfrac{2Q}{4Qt+\varphi} \leq 0 \\ \dfrac{\partial P_i}{\partial t} = 2\beta^2 \left[ \dfrac{2b^2(t)}{\varepsilon_m} + \dfrac{\varepsilon_t}{\varepsilon_m} + \dfrac{\varepsilon_m^2}{\varepsilon_t^2} - \dfrac{\varepsilon_m}{\varepsilon_t} - 1 \right] + \dfrac{8Q^2}{(4Qt+\varphi)^2} \geq 0 \\ \Delta_i I = -\dfrac{b_0^2}{\varepsilon_m}\left(1-e^{-2\beta t}\right) - \dfrac{1}{2}\left(e^{-2\beta t}-1\right) - \dfrac{\varepsilon_0}{2\varepsilon_m}\left(1-e^{-2\beta t}\right) \\ \quad -\dfrac{1}{2}\ln\dfrac{\varepsilon_m + (\varepsilon_0-\varepsilon_m)e^{-2\beta t}}{\varepsilon_0} - \dfrac{1}{2}\ln\dfrac{4Qt+\varphi}{\varphi} \leq 0 \end{cases} \qquad (74)$$

It is not difficult to see from expressions (71)—(74) that for this dynamic system the information entropy production rate $P > 0$, its temporal change rate $\partial P/\partial t < 0$ and information entropy production $\Delta_i S = 0$ at initial state $t = 0$: the information dissipation rate $P_i < 0$, its temporal change rate $\partial P_i/\partial t > 0$ and information dissipation $\Delta_i I = 0$ at initial state $t = 0$. The information entropy's $P = 0$, $\partial P/\partial t = 0$ and $\Delta_i S > 0$ at final state $t = t_f$; the information's $P_i = 0$ $\partial P_i/\partial t = 0$ and $\Delta_i I < 0$ at final state $t = t_f$. The information entropy's $P > 0$, $\partial P/\partial t < 0$ and $\Delta_i S > 0$, the information's $P_i < 0$, $\partial P_i/\partial t > 0$ and $\Delta_i I < 0$ for any other time ($0 < t < t_f$).

Substituting expressions (68)(70) into (44) we obtain the space information flow density of this dynamic system at time $t$ through position $x$ as follows

$$J_x(t) = vI_x(t) - Q\dfrac{\partial I_x}{\partial x}$$
$$= \left\{ \left[v + \dfrac{2Q(x-vt)}{\varphi+4Qt}\right]\left[\dfrac{b_t^2}{\varepsilon_m} + \dfrac{1}{2}\left(\dfrac{\varepsilon_t}{\varepsilon_m} - \ln\dfrac{e\varepsilon_t}{\varepsilon_m}\right) + \dfrac{1}{2}\ln\dfrac{(\varphi+4Qt_f)}{\varphi+4Qt} - \dfrac{(x-vt)^2}{\varphi+4Qt}\right] + \dfrac{2Q(x-vt)}{\varphi+4Qt} \right\} p(x,t)$$

(75)

The difference between the information flow density $J_0(0)$ at time $t = 0$ through position $x = 0$ and the information flow density $J_l(\tau)$ at time $t = \tau$ through position $x = l$ is

$$J_0(0) - J_l(\tau) = P_l(\tau) = v\left[ \dfrac{p_{00}b_0^2 - p_{l\tau}b_\tau^2}{\varepsilon_m} + \dfrac{1}{2}\left( \dfrac{p_{00}\varepsilon_0 - p_{l\tau}\varepsilon_\tau}{\varepsilon_m} - p_{00}\ln\dfrac{e\varepsilon_0}{\varepsilon_m} + p_{l\tau}\ln\dfrac{e\varepsilon_\tau}{\varepsilon_m}\right) \right.$$

$$\left. + \dfrac{p_{00}}{2}\ln\left(\dfrac{\varphi+4Qt_f}{\varphi}\right) - \dfrac{p_{l\tau}}{2}\ln\left(\dfrac{\varphi+4Qt_f}{\varphi+4Q\tau}\right) \right] > 0 \qquad (76)$$

where $\tau = l/v$, $p_{00} = p(x=0, t=0) = 2(\pi\varphi)^{-1/2}$, $p_{l\tau} = p(x=l, t=\tau) = 2(\pi\varphi+4\pi Q\tau)^{-1/2}$, $P_l(\tau)$ is the cumulative dissipation of information flow density of this system during in transmission process. Expression(76) is just the change expression of information flow density of the special dynamic system described by (66) in transmission process. The change of $J_l(\tau)$ and $P_l(\tau)$ with $\tau$ and $Q$ will be discuss in following expression(84).

## 7.2 Molecular motor

Molecular motor plays important role in life process. It can convert chemical energy into mechanical energy effectively and directly. The directed motion of molecular motor performs in the absence of any macroscopic external force. What is the mechanism of this directed motion? It has stimulated interest of many biologists and physists, and many different models were proposed. Here we use a periodically rocked model[20, 21] of Brownian motor to calculate the information entropy production rate and information dissipation rate of the molecular motor. In this model, molecular motor are regarded as a Brownian particle, which are driven by three forces. That is: the spatial unsymmetrical periodical potential、the temporal periodical force and a Gaussian white thermal noise. The two formers originate from the motor system, the third comes from environment. If $a$ denotes the state of molecular motor, then we have

$$\dot{a} = -\frac{\partial}{\partial a}[U(a) - aV(t)] + \eta(t) = K(a) + \eta(t) \tag{77}$$

We take the spatial unsymmetrical periodical potential[21] $U(a) = -\frac{1}{2\pi}\left[\sin(2\pi a) + \frac{1}{4}\sin(4\pi a)\right]$, temporal periodical force $V(t) = A\sin(\omega t)$, and $\eta(t)$ is a Gaussian white noise. As we hope that the spatial average of $\frac{\partial}{\partial a}U(a)$ as well as the time average of $V(t)$ are zero. Because $V(t)$ is time-dependent, the Fokker-Planck equation being equivalent to the Langevin equation (45) does not exist stationary solution. For this reason we may take $\omega \ll 1$, so that $V(t)$ changes with $t$ very slowly. Thus the system exist quasi-stationary solution, and its probability current can be approximately expressed by [21]

$$J = D\left\{\left[1 - \exp\left(-\frac{LA}{D}\sin(\omega t)\right)\right]^{-1}\int_0^L da\int_0^L da'\exp[\phi(a,t) - \phi(a',t)]\right.$$

$$\left. - \int_0^L da\int_0^a da'\exp[\phi(a',t) - \phi(a,t)]\right\}^{-1} \tag{78}$$

where $\phi(a,t) = [U(a) - aA\sin(\omega t)]/D$, spatial period L=1.

The probability density of quasi-stationary state

$$p_{st}(a) = \frac{J}{D}\exp[-\phi(a,t)]\left\{\left[1 - \exp\left(-\frac{LA}{D}\sin(\omega t)\right)\right]^{-1}\int_0^L \exp[\phi(a,t)]da\right.$$

$$\left. - \int_0^a \exp[\phi(a',t)]da'\right\} \tag{79}$$

Substituting $V(t)$、$\phi(L)$、$J$ and $p_{st}(L) = p_{st}(0)$ into formulas(55)(56), we obtain the information entropy production rate and information dissipation rate for molecular motor in quasi-stationary state

$$P = \frac{d_iS}{dt} = \frac{JLA\sin(\omega t)}{D} \tag{80}$$

$$P_i = \frac{d_iI}{dt} = -\frac{JLA\sin(\omega t)}{D} \tag{81}$$

Expressions (80)(81) show that the entropy production rate $P$ and information dissipation rate $P_i$ are proportional to $J/D$ and $V(t) = A\sin(\omega t)$. Because $J$ is also changes with $V(t)$ and $D$ complicatedly, $P$ and $P_i$ are a complicated nonlinear function of $V(t)$ and $D$.

From the information entropy production rate $P > 0$ and information dissipation rate $P_i < 0$, we can conclude that the efficiency of molecular motor is still smaller than 100% though it is very high. That is, the second law of thermodynamics is also universal for molecular motor.

### 7.3 Gaussian channel

In present Shannon static information theory, the Gaussian channel is typical example. Both its symbol and noise obey Gaussian distribution. Its mutual information and channel capacity are[1-3]

$$H(A;B) = \frac{1}{2}\ln\left(1 + \frac{\sigma_a^2}{\sigma_n^2}\right) \tag{82}$$

$$C = \frac{1}{2}\ln\left(1 + \frac{\sigma_a^2}{\sigma_n^2}\right) \tag{83}$$

where the $\sigma_a^2$ and $\sigma_n^2$ are the variances of the input symbol and noise, i.e. their average powers. Formula (83) just is the famous Shannon formula of channel capacity.

Now let us give the dynamic mutual information and the dynamic information capacity of a Gaussian channel through which dynamic symbol is transmitted. As mentioned in expressions (44)(75), for dynamic Gaussian symbol transmitted through spatial Gaussian channel, the proportion of information flow density $J_l(\tau)$ at time $t=\tau$ transmitted to output terminal $x=l$ to information flow density $J_0(0)$ at time $t=0$ transmitted from input terminal $x=0$ is

$$\delta = \frac{J_l(\tau)}{J_0(0)} = \frac{\left[\frac{b_\tau^2}{\varepsilon_m} + \frac{1}{2}\left(\frac{\varepsilon_\tau}{\varepsilon_m} - \ln\frac{e\varepsilon_\tau}{\varepsilon_m}\right) + \frac{1}{2}\ln\frac{\varphi + 4Qt_f}{\varphi + 4Q\tau}\right](\pi\varphi)^{1/2}}{\left[\frac{b_0^2}{\varepsilon_m} + \frac{1}{2}\left(\frac{\varepsilon_0}{\varepsilon_m} - \ln\frac{e\varepsilon_0}{\varepsilon_m}\right) + \frac{1}{2}\ln\frac{\varphi + 4Qt_f}{\varphi}\right](\pi\varphi + 4\pi Q\tau)^{1/2}} \tag{84}$$

According the expressions (84)(62), for a Gausian channel, if the information transmitted from time $t=0$ and input terminal $x=0$ is

$$H(A_0(0)) = \frac{1}{2}\ln(2\pi e\sigma_a^2) \tag{85}$$

then at time $t=\tau$ and output terminal $x=l$ it reduces to

$$H(A_l(\tau)) = \delta\, H(A_0(0)) = \frac{\delta}{2}\ln(2\pi e\sigma_a^2) = \frac{1}{2}\ln(2\pi e\sigma_{a_l}^2) \qquad (86)$$

where
$$\sigma_{a_l}^2 = (2\pi e)^{\delta-1}\sigma_a^{2\delta} \qquad (87)$$

The expression (86) is still a Gaussian symbol, its average power reduces to $\sigma_{al}^2$ from $\sigma_a^2$. Substituting (86) into formulas (63)(65) and using same method for deriving formulas (82)(83), we obtain the dynamic mutual information and the dynamic information capacity of a Gaussian channel as follows

$$I(A_0(0); B_l(\tau)) = \frac{1}{2}\ln\left[1 + \frac{(2\pi e)^{\delta-1}\sigma_a^{2\delta}}{\sigma_n^2}\right] \qquad (88)$$

$$C_d = \frac{1}{2}\ln\left[1 + \frac{(2\pi e)^{\delta-1}\sigma_a^{2\delta}}{\sigma_n^2}\right] \qquad (89)$$

It can be seem from expression (84) that $\delta$ changes with the transmitting time $\tau = l/v$. i.e. the proportion of the channel length $l$ to the transmitting rate $v$ and the space noise strength $Q$. When $\tau = l/v \to 0$. i.e. $l \to 0$ or $v \to \infty$, then $\delta=1$, $J_l(\tau) = J_0(0)$. Under this condition, the cumulative information dissipation in the transmission process $H(Q_l(\tau)) = 0$, $\sigma_{al}^2 = \sigma_a^2$, the dynamic mutual information and the dynamic channel capacity (88)(89) reduce to the static mutual information and the static channel capacity (82)(83). In fact, $\tau = l/v \to 0$ means in physics that the channel length $l$ shortens into one point, the noise of course also concentrates on this point. This is the present static (or zero transmitting time) communication model. When $\tau = l/v \to \infty$. i.e. $l \to \infty$ or $v \to 0$, or $Q \to \infty$, then $\delta = 0$, $J_l(\tau) = 0$. Under this condition, $H(Q_l(\tau)) = H(A_0(0))$, $H(A_l(\tau)) = 0$, all the information of symbol $A_0(0)$ transmitted from input terminal dissipates entirely in transmission process, it cannot receive any information again in output terminal. Under any other general condition, i.e. when $0 < \tau < \infty$ and $0 < Q < \infty$, then $0 < \delta < 1$, $\sigma_{al}^2 < \sigma_a^2$, the dynamic mutual information and the dynamic channel capacity are smaller respectively then the static mutual information and the static channel capacity. All these results just are the modification of the dynamic mutual information and the dynamic channel capacity on the static mutual information and static channel capacity.

Up to now all the results in this paper is based on the dynamic information entropy (7) and the dynamic information (24). If we start from the dynamic information (24a), substituting expression (68) into (44a), we obtain the proportion of the information flow density of the Gaussian channel at the final state to the information flow density at the initial state as follows

$$\delta = \frac{J_l(\tau)}{J_0(0)} = \frac{(\pi\varphi)^{1/2}\ln\left[(\pi e\varepsilon_\tau)(\pi\varphi + 4\pi Q\tau)\right]}{(\pi\varphi + 4\pi Q\tau)^{1/2}\ln\left[(\pi e\varepsilon_0)(\pi\varphi)\right]} \qquad (84a)$$

The physical meaning of expressions (84)(84a) is same, their difference comes from rather different computational starting point. If we use (84a) in place of (84) to calculate the dynamic mutual information and the dynamic channel capacity (88)(89), the qualitative result is same though the quantitative value is rather different.

## 8  Conclusion

Dynamic statistical information theory is an information theory, which deals with the evolution law of dynamic information and dynamic information entropy and its application by means of stochastic process method. Starting from the state variable evolution equation we have derived the nonlinear evolution equations of dynamic information and dynamic information entropy, which describe the evolution law of dynamic information and dynamic information entropy. These two equations are time-reversal asymmetrical. They show that the temporal change rate of dynamic information density originates together from its drift, diffusion and dissipation in coordinate space and state variable space, and the temporal change rate of dynamic information entropy density is caused together by its drift, diffusion and production in coordinate space and state variable space. The expression of information (entropy) flow, the formulas of information dissipation rate and information entropy production rate, and the dynamic mutual information and dynamic channel capacity are derived and presented all from these two evolution equations.

The information flow consists of drift information flow and diffusion information flow. The drift information flow is directional. However, the diffusion information flow always diffuses from a high information density region to all low information density regions. The temporal change rates of the sum of both total and internal information and information entropy in a same dynamic system are equal to zero. The laws of information dissipation and information entropy production show that the dynamic systems always tend spontaneously to the direction of the decrease of the degree of order and the increase of the degree of disorder. These dissipation and production originate from inhomogeneous departure from equilibrium of dynamic system and the internal and external noises. Dynamic mutual information and dynamic channel capacity reflect the dynamic dissipation character in transmission process and change into their maximum—the present static mutual information and static channel capacity under the limit case when the proportion of channel length to signal transmission rate approaches to zero.

As examples of application of the above theoretical formulation, the information dissipation and information entropy production as well as their time rate for two actual dynamic topics are investigate, and the dynamic mutual information and dynamic channel capacity of a Gaussian channel are presented.

Thus, dynamic statistical information theory has accomplished a unified explanation on the evolution law of dynamic information and dynamic information entropy and its application. The present static statistical information theory may be regarded as its a special part of time-space-independent process.

## References

[1] Shannon C E. A mathematical theory of communication [J]. Bell Sys Tech J, 1948, 27: 379-433.

[2] Cover T M, Thomas J A. Elements of information theory [M]. New York: John Wiley & Sons, 1991.

[3] Chang J. Elements of information theory [M]. Beijing: Tsinghua University Press, 1993. (in Chinese)

[4] Weber B H, Depew D J, Smith J D. Entropy, information and evolution [M]. Cambridge: The MIT Press, 1988.

[5] Brillouin L. Science and information theory [M]. New York: Academic Press, 1962.

[6] Kapur J N, Kesavan H K. Entropy optimization principle with application [M]. San Diego: The MIT Press, 1988.

[7] Zurek W H. Ed. Complexity, entropy and the physics of information [M]. Redwood City, California: Addison-Wesley, 1990.

[8] Haken H. Information and self-organization [M]. Berlin: Springer-Verlag, 1988.

[9] Ingarden H S, Kossakowski A, Ohya M. Information dynamics and open system [M]. Dordrecht: Kluwer Academic Publishers, 1997.

[10] Atmanspacher H, Scheingraber H. Information dynamics [M]. New York: Plenum Press, 1991.

[11] Barwise J. Information flow [M]. Cambridge: Cambridge University Press, 1997.

[12] Weenig W H. Information diffusion and persuation in communication networks [M]. Leiden: Leiden University, 1991.

[13] Xing X S. On the fundamental equation of nonequilibrium statistical physics [J]. Int J Mod Phys B, 1998, 12(20): 2005-2029.

[14] Xing X.S. New progress in the principle of nonequilibrium statistical physics [J]. Chinese Science Bulletin, 2001,46(6): 448-454

[15] Xing X.S. On the formula for entropy production rate [J]. Acta Physica Sinica, 2003, 52(12): 2969-2976. (in Chinese)

[16] Wu D J, Cai L, Ke S Z. Bistable kinetic model driven by correlated noise [J]. Phys Rev E, 1994, 50: 2496-2502.

[17] Huang C F. Principle of information diffusion [J]. Fuzzy Sets and System, 1997, 91(1): 69-90.

[18] Frieden B R. Physics from Fisher information [M]. Cambridge: Cambridge University

    Press, 1998.

[19] Haken H. Synergetics [M]. Berlin: SpringerVerlag, 1983.

[20] Magnasco M O. Forced thermal ratchets [J]. Phys Rev Lett, 1993, 71(10): 1477-1480.

[21] Bartussek R, Hanggi P, Kisser J G. Periodically rocked thermal ratchets [J]. Europhys lett, 1994, 28: 459-464.